\documentclass{article}[12pt]

\usepackage[english]{babel}

\usepackage[letterpaper,top=2cm,bottom=2cm,left=3cm,right=3cm,marginparwidth=1.75cm]{geometry}

\usepackage{amsmath}
\usepackage{amsthm}
\usepackage{graphicx}
\usepackage{float}
\usepackage[colorlinks=true, allcolors=blue]{hyperref}
\usepackage{natbib}

\title{A Levered ETF Anomaly Explained}
\author{Stephen W. Bianchi\footnote{University of California, Berkeley}\hspace{1mm} and Lisa R.\thinspace Goldberg\footnote{BlackRock, Inc. and University of California, Berkeley}\thanks{The authors are grateful to Robert M. Anderson for substantial contributions to this article and to Ananth Madhavan for educating us about levered and inverse levered ETFs. Thanks to Hank Bessembinder, Patrick Geddes, Ron Kahn, Alec Kercheval, Petter Kolm and Ken Ribet for insightful comments on early drafts.}}
\date{\today}
\begin{document}

\maketitle

\begin{abstract}
Counterintuitively, the S\&P 500 Index rose between January 1, 2022, and December 29, 2023, while exchange-traded funds (ETFs) seeking to deliver $2\times$ and $3\times$ daily returns of the index delivered substantially negative returns. Roughly two-thirds of the difference between the returns of the index and the levered ETFs can be attributed to compounding and volatility. The remaining difference is explained by the covariance between the ETFs’ deviations from constant leverage and the index’s return.
\end{abstract}

\section{Investing in the US market}

The US equity market has enjoyed double digit returns  for decades.  Many investors would be pleased  had their portfolios grown like that, but the US market isn't an investable security. Its growth is indicated by the performance of benchmarks like the S\&P 500 Index,  which are  weighted averages of prices of many investable securities. Once, holding a portfolio that effectively mimicked performance of the US market was impractical.

 The opportunity  for retail investors to obtain direct exposure to the growth of the US market arrived on December 31, 1975, when Vanguard founder Jack Bogle introduced the First Index Investment Trust. It was constructed to closely track the S\&P 500 Index with the aim of ``giving investors their fair share of market returns" at low cost. Maligned early on as ``un-American" and nicknamed ``Bogle's Folly," the First Index Investment Trust was launched with \$11 million and made \$17 million in its first five years of operation.  It crossed  \$100 billion in assets under management in November, 1999. 

Since Bogle's Folly, index-tracking funds and the practice of passive investing have grown both in  popularity and their impact on financial markets.  This trend was accelerated  by the introduction in 1999 of the exchange traded fund (ETF), a security designed to mimic the return stream of an index. ETFs tend to be relatively low-cost and tax-efficient, and they are distinguished by their creation redemption mechanism. While the largest ETFs, such as the Vanguard S\&P 500 ETF (VOO), aim to closely track capitalization-weighted large-cap equity benchmarks, many have more exotic missions.

We focus on ProShares Ultra S\&P500 (SSO) and 
ProShares UltraPro S\&P500 (UPRO), ETFs that seek to deliver double and triple the return of the S\&P 500 Index on a daily basis.  The appeal seems obvious: if  index returns are attractive, doubling or tripling those returns could be even better. As is well known, however, a wide range of disparities between the  performance of an index and a daily levered ETF on that index may occur over a long horizon.  Factors that determine the disparity include  index return and volatility as well as the length of the investment horizon. These factors are discussed in SSO and UPRO's prospectuses.

We report on a two-year period where the returns of  the S\&P 500 Index and its tracker VOO were slightly positive, while the returns of SSO and UPRO were substantially negative.  This can be attributed directionally to a mathematical approximation of geometric return in terms of arithmetic return and volatility.  As we show, that approximation leaves a substantial error.  To explain this, consider that a
levered ETF may target a multiple of index return calculated from net asset value (NAV) of individual securities. However, investors realize returns based on market prices. Our analysis, therefore, uses ETF price based returns, which incorporate both portfolio leverage and discrepancies between price and NAV. As a consequence of this and perhaps other considerations, price returns of a levered ETF deviate from a constant target, especially when index returns are small in magnitude. The timing of these deviations relative to index returns explains  the residual error.

In Section~\ref{sec:daily}  we look at the daily performance of the S\&P 500 Index along with VOO, SSO and UPRO from January 2, 2022 to December 29, 2023. Section~\ref{sec:long}  reports performance over the full two-year period. In Section~\ref{sec:vol}, we attempt to reconcile the mildly positive performance of the index over the full two-year period with the substantially  negative performance of SSO and UPRO  using standard factors:  average return, volatility and leverage, and find a substantial gap.  In Section~\ref{sec:lev}, we complete the reconciliation by considering the covariance between small changes in leverage of the ETFs and returns to the index. Section~\ref{sec:future} provides guidance on what is needed to forecast future returns of levered ETFs.

\section{ETFs versus the index: a daily perspective}\label{sec:daily}

Figure~\ref{fig:scatter} shows scatter plots of  daily returns of VOO, SSO and UPRO versus  S\&P 500 Index returns over the period  from January 3, 2022 to December 29, 2023.  The plots  show tight  linear relationships at the intended scales.

\begin{figure}[H]
    \centering
    \includegraphics[trim=0 0 10cm 0,clip,width=0.85\linewidth]{images/figure1_voo.png}
    \caption{Daily returns (\%) of VOO, SSO and UPRO versus the S\&P 500 Index. January 3, 2022--December 29, 2023.  Returns of VOO, SSO and UPRO are approximately $1\times$, $2\times$ and $3\times$ the returns of the S\&P 500 Index. }
    \label{fig:scatter}
\end{figure}

\section{ETFs versus the index: A long-horizon perspective}\label{sec:long}

\begin{figure}[H]
    \centering
    \includegraphics[width=0.7\linewidth]{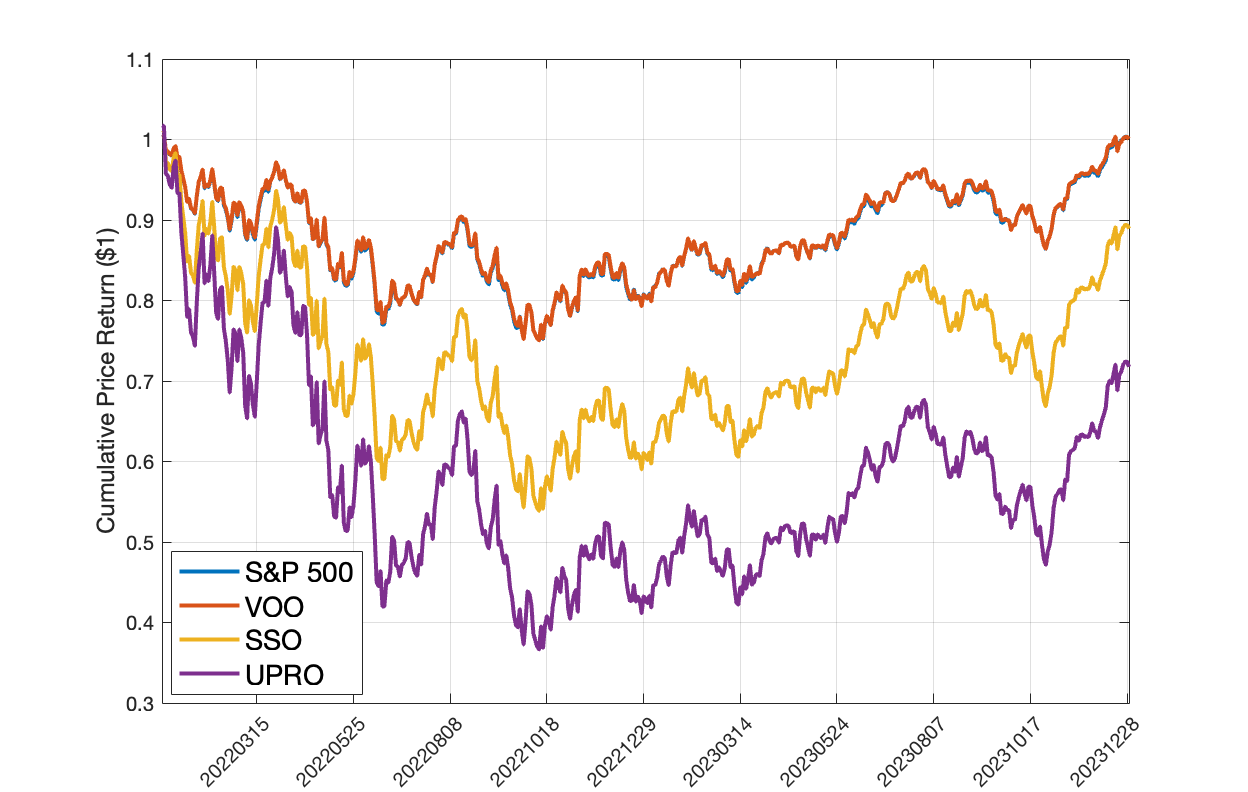}
    \caption{Cumulative returns  of the S\&P 500 Index, VOO, SSO and UPRO. January 3, 2022--December 29, 2023. Returns of the index and VOO were mildly positive while returns of SSO and UPRO were substantially negative.}
    \label{fig:cumret}
\end{figure}

 The tight relationships between daily returns of levered ETFs and the underlying index shown in Figure~\ref{fig:scatter} need not be present at longer horizons.  Between January 3, 2022 and December 29, 2023, the S\&P 500 Index was up slightly, earning a cumulative return of 0.076\%, while SSO and UPRO were down, earning $-11.09$\% and $-28.24$\%.  Summary statistics on the index and three ETFs between January 3, 2022 and December 29, 2023 are in Table~\ref{table:stats} and their cumulative returns are shown in Figure~\ref{fig:cumret}.

\begin{table}[H]
\begin{center}
\begin{tabular}{| l |  r  r  r  r |}
\hline
 & S\&P 500 & VOO  & SSO  & UPRO  \\
\hline
Cumulative Return & 0.076 & 0.053 & -11.088 & -28.238 \\
Geometric Return & 0.038 & 0.026 & -5.707 & -15.288 \\
Arithmetic Return & 1.917 & 1.897 & 1.655 & 0.309 \\
Volatility & 19.40 & 19.36 & 38.77 & 58.01 \\
Risk-Adjusted Return & 0.099 & 0.098 & 0.043 & 0.005 \\
Days in Sample & 501 & 501 & 501 & 501 \\
\hline
\end{tabular}
\end{center}
\caption{Summary statistics. Returns and volatility are in annualized \%. January 3, 2022--December 29, 2023.}
\label{table:stats}
\end{table}

%\begin{table}[H]
%    \centering
%    \includegraphics[width=0.7\linewidth]{images/stats.png}
%    \caption{Summary statistics. January 3, 2022--December 29, 2023.}
%    \label{table:stats}
%\end{table}

How is it possible that the return of a security that effectively doubled the S\&P 500 Index return on a daily basis was negative while the index appreciated in value?  As is well known,  the complex relationship between arithmetic and geometric return is a contributing factor. Over long horizons, securities  grow at the {\it geometric} average daily return.  While it can never be larger than an arithmetic average, a geometric average is often smaller than its arithmetic counterpart, sometimes by a wide margin. As we illustrate in the next section, the way in which a geometric average tends to increase with arithmetic average and to decrease with volatility can be made mathematically precise.

\section{The impact of volatility}\label{sec:vol}

Consider an investment in a portfolio that is worth \$1 on day $t=0$ and grows to a value $V$ on day $t=T$.
Neglecting dividends and market frictions, the arithmetic return on day $t$ is given by
$$R_t = \frac{P_{t+1} - P_t}{P_t},$$
where $P_t$ and $P_{t+1}$ are the prices on days $t$ and $t+1$. The value at period end is given by
$${\rm Value} = \frac{P_T}{P_0} = \prod_{t= 0}^{T-1} \frac{P_{t+1}}{P_t} = \prod_{t= 0}^{T-1}  (1 + R_t)$$
and the daily geometric return is given by
$$G = {\rm Value}^{1/T} - 1 = \left[\prod_{t= 0}^{T-1}  (1 + R_t)\right]^{1/T} -1.$$

Thus, long-horizon performance is completely determined by daily geometric return $G$,  which is bounded above by daily arithmetic return,
$$E = \frac{1}{T} \sum_{t = 0}^{T-1} R_t.$$
% Suppose two investment strategies 
% %deliver returns $r$ and $r'$, with 
% have final values $V$ and $V'$ at the investment horizon of $T$ days.  
%  $V$ and $V'$ always have the same ordering as $G$ and $G'$, but it is entirely possible that the ordering of $E$ and $E'$ is the opposite of $V$ and $V'$.  The gap between $G$ and $E$ is driven by by the volatility of investment returns.  
Arithmetic return, which may be more intuitive than geometric return, need not align with long-term performance. This disconnect can be explained by the relationship between $G$ and $E$, which depends on return volatility 
$V$:
\begin{equation}
G \sim \left(1+E\right)e^{-\frac{V^2}{2}}-1,\label{formula: APR empirical simple}\end{equation}
   which is derived in Appendix \ref{section:Geometric_Return} using a second order Taylor approximation.

\begin{figure}[H]
    \centering
\includegraphics[width=0.7\linewidth]{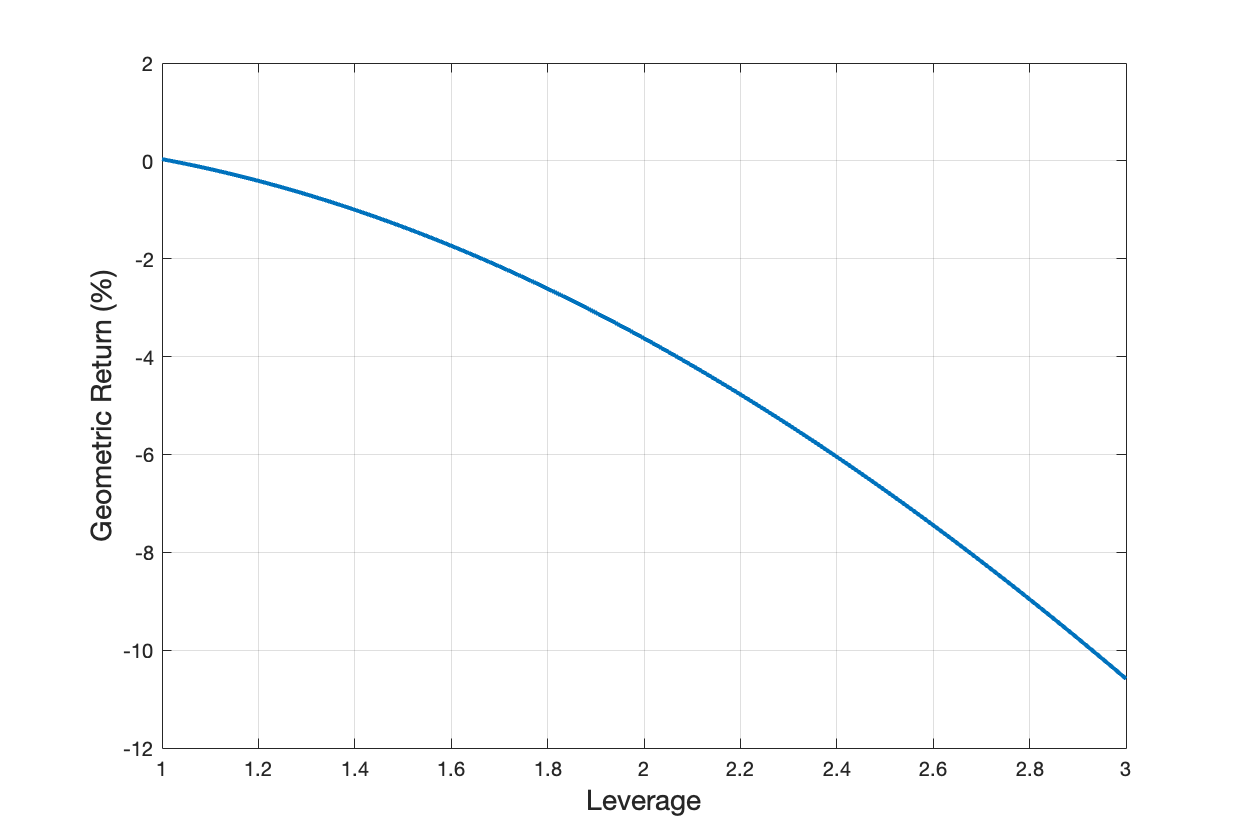}
    \caption{Dependence of annualized geometric return on leverage based on the daily arithmetic return (mean) and volatility of the S\&P 500 Index, January 3, 2022--December 29, 2023, annualized return 1.92\% and annualized volatility 19.40\%.}
    \label{fig:g-lev}
\end{figure}

An investor with estimates of arithmetic return $E$ and volatility $V$ of the S\&P 500 Index might use formula~(\ref{formula: APR empirical simple}) to estimate the geometric return $G$ of the unlevered ETF VOO and levered ETFs SSO and UPRO.  They can obtain an estimate of  $G$ for VOO using estimates of $E$ and $V$ directly, and for SSO and UPRO by assuming ideal constant leverage.  This amounts to scaling $E$ and $V$ of the S\& P 500 Index  $2\times$  and $3\times$. The chart in Figure~\ref{fig:scatter} suggests this may be a reasonable approach, and we tested it with daily data from  January 3, 2022 to December 29, 2023. The results, labelled ``Ideal Constant Leverage Estimate," are in Table~\ref{table:ourest}.

\begin{table}[H]
\begin{center}
\begin{tabular}{| l | r | r | r | r |}
\hline
 & S\&P 500 & VOO & SSO  & UPRO  \\
\hline
Geometric Return & 0.038 & 0.026 & -5.707 & -15.288 \\
Ideal Constant Leverage Estimate & 0.035 & 0.035 & -3.625 & -10.580 \\
\hline
\end{tabular}
\end{center}
\caption{\label{table:ourest} Annualized geometric returns of the S\&P 500 Index. Ideal Constant Leverage Estimates are obtained by  scaling the arithmetic mean and volatility of the index return  by $1\times$, $2\times$ and $3\times$  for VOO, SSO and UPRO,  applying Formula~(\ref{formula: APR empirical simple}). January 3, 2022--December 29, 2023. Though estimates for SSO and UPRO geometric returns are negative while the S\&P 500  Index increased in value, these negative estimates exceed actual performance. }
\end{table}
 The estimates for geometric returns of SSO and UPRO over the period January 3, 2022 to December 29, 2023 are substantially negative while the S\&P 500 Index increased in value.  These estimates exceed the actual performance of the levered ETFs by wide margins. As we illustrate below, the discrepancy arises from the failure of SSO and UPRO to maintain constant $2\times$ and $3\times$ leverage on a daily basis.

\section{The impact of timing}\label{sec:lev}

\begin{figure}[H]
    \centering
    \includegraphics[width=0.9\linewidth]{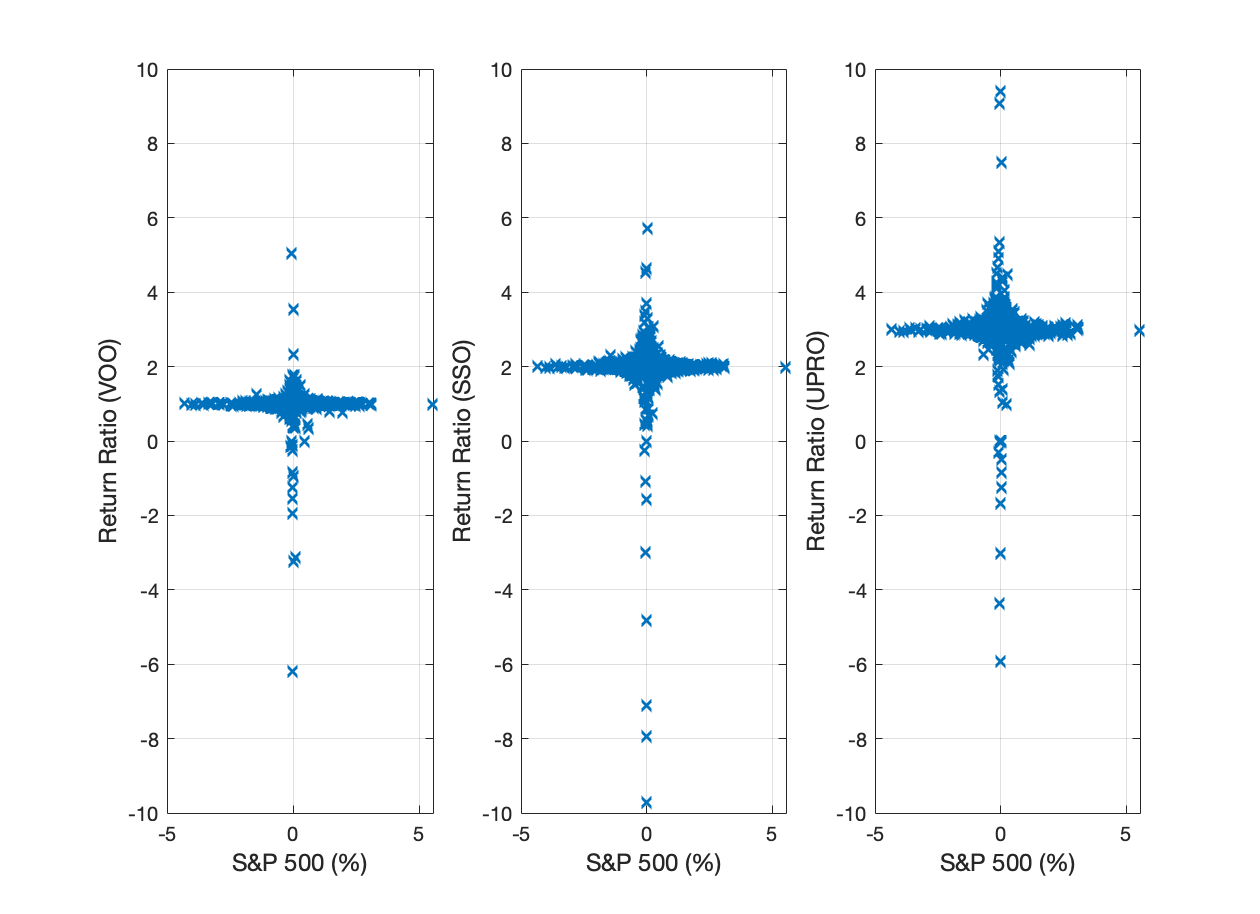}
    \caption{\label{fig:rr} Daily Return Ratios for VOO, SSO and UPRO.  January 3, 2022--December 29, 2023. Deviations from the horizontal lines $y = 1$ for VOO, $y = 2$  for SSO and $y=3$ for UPRO indicate that ETF returns are not perfectly levered index returns.}
    \label{figure:cloudVOO}
\end{figure}

The ETFs VOO, SSO and UPRO aim to deliver daily returns of the S\&P 500 Index scaled $1 \times$, $2\times$ and $3\times$.  Scatter plots of daily returns of these ETFs against the index  shown in Figure~\ref{fig:scatter} indicate that the ETFs largely deliver on their promise.
However, a different visualization of the same data highlights discrepancies that are hard to discern in Figure~\ref{fig:scatter}. In Figure~\ref{fig:rr}, we focus on errors rather than values: we replace the dependent variables in the scatter plots in Figure~\ref{fig:scatter} with ratios of returns of the ETFs to index returns.  The return ratios may be viewed as an approximation to leverage.
Since there is, in the usual sense,  no leverage in VOO---its return ratios relative to the S\&P 500 Index may be viewed as approximations of 1.0---we see that return ratios incorporate errors in addition to leverage.  A possible source of the deviation of ETF return ratios from target is the discrepancy between net asset value, which is the market value of the underlying holdings, and the price of the ETF.\footnote{We thank Hank Bessembinder for this insight.}
The return ratio is constructed directly from observed daily returns and, as we show below, can explain observed components of long-horizon returns.  It is not intended to recover portfolio or target leverage, but rather to measure realized effective leverage from the perspective of investors.

The charts in Figure~\ref{fig:rr} show that daily return ratios form a cloud of values around the horizontal lines $y = 1.0$ for VOO, $y = 2.0$ for SSO, and around $y =3.0$ for UPRO.  Deviations from those lines indicate that ETF price returns are not ideally levered index  price returns.

The clouds are thickest near the $y$-axis,  corresponding to days on which the index return is close to zero.  This is not surprising;  an error in the numerator of a ratio makes a bigger difference when the denominator is small.   What is surprising is the profound impact the deviations in return ratio from target have on performance. As we show next, they explain the substantial differences between actual geometric returns of SSO and UPRO and Ideal Leverage  Estimates shown in Table~\ref{table:ourest}.

On an annualized basis, the average daily return of the S\&P 500 Index over the period from January 3, 2022 to December 29 was 1.92\%.   If SSO were constantly levered $2\times$ on a daily basis, its annualized average daily return would have been $2 \times 1.92\% = 3.84\%$. If, instead, we scaled the index return by the return ratio of 2.03 of SSO, the annualized  return of SSO would have been $2.03 \times 1.92\% = 3.90\%$. The actual average daily return of SSO was 1.65\%.

 The difference is explained by  comovement of index return and  the return ratio. Suppose the daily returns of the S\&P~500 Index and an ETF are denoted $R_t$ and  $R^{\rm ETF}_t$.  Let  $\lambda^{\rm ETF}_t = R^{\rm ETF}/R_t$ be the daily  return ratio.  Then 
\begin{align} E(R^{\rm ETF}) &= \frac{1}{T} \sum_{t = 0}^{T-1} R^{\rm ETF}_t \nonumber\\
  &= \frac{1}{T} \sum_{t = 0}^{T-1}\lambda^{\rm ETF}_t R_t \nonumber \\ &= E(\lambda^{\rm ETF})E(R) + {\rm Cov} (\lambda^{\rm ETF}, R).\label{eq:cov}\end{align}
Rearranging  Formula~(\ref{eq:cov}) gives an expression for the covariance between the return ratio and the return to the S\&P 500 Index:
\begin{align}\label{eq:covSS}
    {\rm Cov} (\lambda^{\rm ETF}, R) =E(R^{\rm ETF}) - E(\lambda^{\rm ETF})E(R).
\end{align}
Table~\ref{table:mean} reports annualized arithmetic returns.  For SSO and UPRO, the actual returns are noticeably smaller than the constant leverage estimates, leading to substantial discrepancies given by the covariance term in Formula~\ref{eq:covSS}.

\begin{table}[H]
\begin{center}
\begin{tabular}{| l | r | r | r | r |}
\hline
 & S\&P 500 & VOO & SSO  & UPRO  \\
\hline
Arithmetic Return & 1.917 & 1.897 & 1.655 & 0.309 \\
Average Return Ratio & 1.000 & 1.028 & 2.033 & 3.106 \\
Volatility of Return Ratio & 0.000 & 20.213 & 38.926 & 48.362 \\
\hline\hline
Ideal Constant Leverage Estimate Arithmetic Return & 1.917 & 1.917 & 3.833 & 5.750 \\
Empirical Constant Leverage Estimate Arithmetic Return & 1.917 & 1.970 & 3.897 & 5.952 \\
\hline\hline
Covariance($\lambda^{\rm ETF}, R$) & 0.000 & -0.073 & -2.242 & -5.643 \\
\hline
\end{tabular}
\caption{\label{table:mean} Annualized arithmetic daily returns, average and standard deviation return ratios, constant leverage annualized return estimates  and annualized discrepancies.  Ideal and Empirical Constant Leverage Estimates are obtained by scaling the annualized mean daily return of the S\&P 500 Index by ideal leverage (1.0 for the S\&P 500 Index, 1.0 for VOO, 2.0 for SSO, 3.0 for UPRO)  and average empirical leverage (1.0 for the S\&P 500 Index, 1.028 for VOO, 2.033 for SSO, 3.106 for UPRO).  January 3, 2022--December 29.  A negative value for the covariance between return ratio and S\&P 500 Index return indicates a negative association between ETF return ratio and index return }
\end{center}
\end{table}

% \begin{table}[H]
% \begin{center}
% \begin{tabular}{ l c c c c}
% &S\&P 500 & VOO & SSO & UPRO \\ 
% Arithmetic Return &1.92& 1.93& 1.65 & 0.31  \\  
% Average Return Ratio  &1.00&1.15&2.03 & 3.10\\
% Standard Deviation Return Ratio &0.00 & 3.48 & 2.46 & 3.06\\ \\
% \hline \\
% Ideal Constant Leverage Estimated Arithmetic Return  &1.92& 1.92 &3.84 & 5.76 \\
% Empirical Constant Leverage Estimated Arithmetic Return &1.92 & 2.21&3.90 & 5.96\\ \\ \hline \\
% Discrepancy (  ${\rm D}^{\rm ETF}$) &0.00&-2.78& -2.25 & -5.65
% \end{tabular}
% \caption{\label{table:mean} Annualized arithmetic daily returns, average and standard deviation return ratios, constant leverage annualized return estimates  and annualized discrepancies.  Ideal and Empirical Constant Leverage Estimates are obtained by scaling the annualized mean daily return of the S\&P 500 Index by ideal leverage (1.0 for the S\&P 500 Index, 1.0 for VOO, 2.0 for SSO, 3.0 for UPRO)  and average empirical leverage (1.0 for the S\&P 500 Index, 1.15 for VOO, 2.03 for SSO, 3.10 for UPRO). Discrepancy is obtained by subtracting the Empirical Constant Leverage Estimated  Arithmetic Return from the (actual) Arithmetic Return. January 3, 2022--December 29.  A negative value for Discrepancy indicates a negative association between return ratio and return in the subsequent period. }
% \end{center}
% \end{table}

The covariance term can be viewed as a timing effect.\footnote{We thank Petter Kolm for this framing.} A negative sign on the covariances between the ETF return ratio and the index return indicates that deviations from target leverage tended to increase when the index return decreased, and vice versa.   This timing effect is present in empirical data but neglected when we estimate the average daily return of SSO, UPRO, or any other levered ETF by simply scaling the return of the S\&P~500 Index by a constant.  A summary of this analysis is in Table~\ref{table:mean}.

We repeat Table~\ref{table:ourest} with an additional line, where we use the empirical arithmetic returns and volatilities of the ETFs.
There are substantial errors in constant leverage estimates of geometric return. The empirical estimate, which takes account of the covariance between  is more accurate.

\begin{table}[H]
\begin{center}
\begin{tabular}{| l | r | r | r | r |}
\hline
 & S\&P 500 & VOO & SSO  & UPRO  \\
\hline
Geometric Return & 0.038 & 0.026 & -5.707 & -15.288 \\
Ideal Constant Leverage Estimate & 0.035 & 0.035 & -3.625 & -10.580 \\
Empirical Constant Leverage Estimate & 0.035 & -0.018 & -3.806 & -11.478 \\
Empirical Estimate & 0.035 & 0.024 & -5.694 & -15.226 \\
\hline
\end{tabular}
\end{center}
\caption{\label{table:ourest2} Annualized geometric returns of the S\&P 500~Index and three ETFs.  Ideal Constant Leverage Estimates are obtained by  scaling the sample mean and volatility of the index return  by $1\times$, $2\times$ and $3\times$ for VOO, SSO and UPRO and  applying Formula~(\ref{formula: APR empirical simple}).  Empirical Constant Leverage Estimates are obtained the same way, except sample meanand volatility of the index return is scaled by empirically observed average return ratio, $1.028\times$, $2.033\times$ and $3.106\times$ for VOO, SSO and UPRO. The Empirical Estimate uses the same mean returns and volatilites of the ETFs in Formula~(\ref{formula: APR empirical simple}). The Empirical Estimate is close to the actual value because it accounts for both the compounding in geometric returns and the covariance between ETF return ratio and index return.}
\end{table}

\section{Forecasting returns to levered ETFs over long horizons}\label{sec:future}
An investor with forward-looking views of  the arithmetic return and volatility of the S\&P 500 Index over a future horizon can use Formula~(\ref{formula: APR empirical simple}) to forecast geometric return of ETFs like VOO, SSO and UPRO, assuming constant leverage. During the period from January 3, 2022 to December 29, 2023, this method qualitatively explained the fact that the S\&P 500 Index had a mildly positive return while the levered ETFs SSO and UPRO had substantially negative returns. However, there were large discrepanies between estimates that relied on the assumption of ideal leverage and empirically observed performance.   Those discrepancies can be explained by the covariance of index return and return ratios of the ETFs. 
% Consequently, an accurate forecast of levered ETF returns over long horizons requires  a forecast of that covariance.
 Within this framework, forecasting long-horizon levered ETF returns requires a view on the covariance term.

Return to a daily-levered ETF over a long horizon is determined by mean return and volatility of the underlying index, target leverage and covariance between index return and return ratio of the ETF, and all of the components can make important contributions.

\appendix

\section{The impact of return ratio outliers}

% Target leverage can be estimated with the ratio of the return to an ETF to index return.
The covariance of index return and ETF return ratio  explained a large component of the two-year returns of SSO and UPRO between January 3, 2022 and December 29, 2023.  Here, we look more closely at this finding.

Figure~\ref{figure:cloudVOO} plots daily return ratios of VOO, SSO and UPRO against daily returns of the S\&P 500 Index between January 3, 2022 and December 29, 2023.  The most extreme outliers occur when the index returns are close to zero.  This is consistent with the fact that an estimated ratio is especially prone to large errors when the denominator is small. To assess the impact of these large outliers, we winsorized the return ratios (with respect to each of the individual return ratio empirical distributions) at the 95th percentile, and redid the analysis in Table~\ref{table:mean}.
According to the results shown in Table~\ref{table:mean2}, winsorization diminished Covariance$(\lambda^{\rm ETF}, R)$ by a small margin.

\begin{table}[H]
\begin{center}
\begin{tabular}{| l | r | r | r | r |}
\hline
{\it Winsorized (95th Percentile)} & S\&P 500 & VOO & SSO  & UPRO  \\
\hline
Arithmetic Return & 1.917 & 1.897 & 1.655 & 0.309 \\
Average Return Ratio & 1.000 & 0.983 & 1.988 & 2.935 \\
Volatility of Return Ratio & 0.000 & 15.550 & 31.437 & 46.408 \\
\hline\hline
Ideal Constant Leverage Estimate Arithmetic Return & 1.917 & 1.917 & 3.833 & 5.750 \\
Empirical Constant Leverage Estimate Arithmetic Return & 1.917 & 1.885 & 3.811 & 5.626 \\\hline\hline
Covariance($\lambda^{\mbox{ETF}}$, $R$) & 0.000 & 0.250 & -2.215 & -5.100 \\
\hline
\end{tabular}
\caption{\label{table:mean2} Annualized arithmetic daily returns, average and standard deviation of return ratios, constant leverage annualized return estimates  and annualized discrepancies. Return ratios for VOO, SSO, and UPRO are winsorized (with respect to each of the individual return ratio empirical distributions) at the 95th percentile. Ideal and Empirical Constant Leverage Estimates are obtained by scaling the annualized mean daily return of the S\&P 500 Index by ideal leverage (1.0 for the S\&P 500 Index, 1.0 for VOO, 2.0 for SSO, 3.0 for UPRO)  and average empirical leverage (1.0 for the S\&P 500 Index, 0.983 for VOO, 1.988 for SSO, 2.935 for UPRO).  January 3, 2022--December 29.  Removal of outliers diminished Covariance$(\lambda^{\rm ETF}, R)$ by a small margin.}
\end{center}
\end{table}

\section{Related Literature}

ETF basics are explained in \cite{madhavan2016} and \cite{lettau2018}.     An analysis of levered and inverse levered ETFs is in \cite{cheng2009}. \cite{avellaneda2010}
and \cite{jarrow2010} extend this analysis to allow for time-varying volatility and positive interest rates.

Possible outcomes of  SSO and UPRO in terms of expected return and risk of the S\&P 500 Index are discussed in \cite{SSO_Prospectus} and \cite{UPRO_Prospectus}.
The derivation in Appendix~\ref{section:Geometric_Return} of Formula~(\ref{formula: APR empirical simple}) for the approximate relationship between geometric return, arithmetic return and volatility  is taken from Appendix A of \cite{anderson2014}. Similar relationships are derived in \cite{booth1992} and \cite{markowitz2012}, to name just two examples. \cite{anderson2012} discusses related leverage anomalies.

\cite{lenkey2024} provides a survey of the literature related to the potential for levered and inverse levered ETFs to affect late-date asset prices.

 \cite{bessembinder2025} analyzes the performance of  single-stock, levered ETFs relative to a simple benchmark. 
\section{Approximating Geometric Return in Terms of Arithmetic Return and Volatility}\label{section:Geometric_Return}

We derive Formula~\ref{formula: APR empirical simple} from Appendix A of \cite{anderson2014}.

\begin{eqnarray} 
\log\left(1+G\right) &=& 
\frac{1}{T} \sum_{t=0}^{T-1} \log \left(1+R_t\right)\nonumber\\
&\sim& \frac{1}{T}\sum_{t=0}^{T-1} \left(R_t - \frac{R_t^2}{2}\right)\label{formula:approximationerror}\\
&=& \frac{1}{T}\sum_{t=0}^{T-1} R_t -\frac{1}{T} \sum_{t=0}^{T-1}\frac{R_t^2}{2}\nonumber\\
&=& E-\frac{V^2+ E^2}{2}\nonumber\\
&\sim& \log \left(1+E\right)e^{- \frac{V^2}{2}}\label{formula:APR empirical}\\
G &\sim& \left(1+E\right)e^{-\frac{V^2}{2}}-1\nonumber\end{eqnarray}
Formulas (\ref{formula:approximationerror}) and (\ref{formula:APR empirical}) approximate the logarithm by its quadratic Taylor polynomial.  When $R_t>0$, the Taylor series for the logarithm is alternating and decreasing in absolute value for $\left|R_t\right|<1$, so the error in the approximation of $\log\left(1+R_t\right)$ in Formula 
(\ref{formula:approximationerror}) is negative and bounded above in magnitude by $\left|R_t\right|^3/3$ for each day $t$.  When $R_t<0$, the error is positive and may be somewhat larger than $\left|R_t\right|^3/3$.  Since the daily returns are both positive and negative, the errors in days with negative returns may substantially offset the errors in days with positive returns, so the errors may tend not to accumulate over time. 
\section{Data}
\medskip\noindent
The results we have presented were based on CRSP daily ETF return data from Janaury 3, 2022 through December 29, 2023. All ETF returns and returns of the S\&P 500 were taken from the CRSP Daily Stock table. The S\&P 500 returns (variable name \emph{sprtrn}) are price returns (i.e. do not include dividends), as are all of the ETF returns (variable name \emph{retx}).

% \medskip \noindent
% VOO is the symbol for the the {\bf State Street SPDR S\&P 500 ETF Trust}. The trust seeks to achieve its investment objective by holding a portfolio of the common stocks that are included in the index, with the weight of each stock in the trust substantially corresponding to the weight of such stock in the index.

 \medskip \noindent
VOO is the symbol for the {\bf Vanguard S\&P 500 ETF}.The fund employs an indexing investment approach designed to track the performance of the Standard \& Poor's 500 Index. The advisor attempts to replicate the target index by investing all, or substantially all, of its assets in the stocks that make up the index, holding each stock in approximately the same proportion as its weighting in the index. The fund is non-diversified.

% \medskip \noindent
% IVV is the symbol for {\bf iShares Core S\&P 500 ETF}. The fund generally will invest at least 80\% of its assets in the component securities of the index and in investments that have economic characteristics that are substantially identical to the component securities of its index and may invest up to 20\% of its assets in certain futures, options and swap contracts, cash and cash equivalents.

\medskip \noindent
SSO is the symbol for {\bf ProShares Ultra S\&P500}. SSO seeks daily investment results, before fees and expenses, that correspond to two times (2x) the daily performance of the S\&P 500. \footnote{\url{https://www.proshares.com/globalassets/proshares/fact-sheet/prosharesfactsheetsso.pdf}}

\medskip \noindent
UPRO is the symbol for {\bf ProShares UltraPro S\&P500}.  UPRO seeks daily investment results, before fees and expenses, that correspond to three times (3x) the daily performance of the S\&P 500. \footnote{\url{https://www.proshares.com/globalassets/proshares/fact-sheet/prosharesfactsheetsso.pdf}}

% \section{How a Levered ETF Works }

% The obvious way to construct a levered strategy is to borrow, but borrowing costs erode both the return and the Sharpe ratio of the levered portfolio, even in the absence of transaction costs associated with rebalancing; see the discussions in \cite{anderson2012} and \cite{anderson2014}.   In practice, borrowing costs can be reduced but not eliminated through the use of derivatives such as futures contracts.\footnote{See \cite{GAGP}.}  However, the use of derivatives  {\color{blue}  Do we know that the issue is derivatives? could it have something to do with timing, or failure to hold all stocks?  Let's not be specific if we're not sure.} prevents the exact realization of the intended replication strategy.

\bibliographystyle{agsm}
\bibliography{sample}
\end{document}